\documentclass[10pt,journal,letterpaper,twocolumn]{IEEEtran}
\usepackage{amsmath,epsfig,algorithmic}
\usepackage[boxed]{algorithm}

\begin{document}

\title{Progressive InSAR Phase Estimation}
\author{Francesco De Zan \thanks{%
DLR (German Aerospace Center), Oberpfaffenhofen,
D-82230 Wessling, Germany (e-mail:francesco.dezan@dlr.de)}}
\maketitle

\begin{abstract}
This paper introduces a novel scheme to progressively estimate interferometric phases from a stack of synthetic aperture radar (SAR) images. The scheme is shown to yield comparable performance to full-covariance algorithms for a realistic decorrelation scenario. The implementation is suited for continuous processing and updating of phase products, without compromising long-term phase accuracy. It also limits the requirements in terms of data transfer between archive and processing facility, a significant issue for  processing large archives of SAR data.
\end{abstract}


\section{Introduction}
The work in \cite{Teba2008} was the first to tackle the issue of optimal phase estimation for SAR interferometry. It proposed a maximum-likelihood algorithm for phase retrieval, assuming knowledge of the underlying temporal covariance. This proposal and many similar other (\cite{DeZan2007}, \cite{Fornaro2015}, \cite{Samiei2016}, \cite{Ansari2018}) are based on the covariance matrix of the acquisitions and therefore suffer from the same limitation: they need access to the entire history of acquisitions. 

We partially tackled this limitation proposing a ``sequential estimation''~\cite{Ansari2017}. In that case there is no need to have access to the entire past, having a compressed version of the past is enough. However that algorithm works in batches, for example ingesting six months or one year of acquisitions at a time. I would like to propose something that incorporates new acquisitions as they come, producing a new phase product for each single-look complex image that becomes available from a SAR mission. This seems to be very valuable for users and therefore a reasonable requirement for an operational phase product.

According to our experience, the various phase triangulation variants cited above suffer from some noisiness in short-term interferograms. We have documented this in \cite{Ansari2018} where the loss of performance in short-term interferograms was quantified in 1-4dB on average. This seems acceptable if the goal is the recovery of long-term deformation, however it is not completely satisfying. The solution seems obviously to limit the algorithms to high-coherent short-term interferograms, but we have also shown \cite{Ansari2020} that using band-limited covariance matrices is very dangerous and is typically subject to phase drifts, as short-term interferograms are often biased. 

Even in the absence of biases, restricting oneself to short-term interferograms might not be a good idea, since the implicit summation of short-time differences to recover long-time series implies an error integration: the resulting series will drift more than necessary if a long-term coherence term is present. The integration of short-time differences is really justified only for a pure auto-regressive coherence model.

My proposal is to design two separate mechanisms to tackle short-term coherence and long-term stability. For the short term, a continuously updated synthetic image will be a reference to estimate the new phases. This {\it running reference} will have reduced noisiness but also represent components of the scatterer which are relatively short lived (e.g. a few months) and sustain short-term coherence.

Periodically, or even at each new acquisition, it will be necessary to adjust (calibrate) the phase of the running reference for long-term stability, to avoid the buildup of phase drifts. To do this I suggest taking snapshots of the running reference and either saving them periodically or accumulating them in a special buffer.

The whole algorithm should have some desirable characteristics: 
\begin{itemize}
\item it should provide a precise and accurate phase, at least comparable to full-covariance approaches
\item it should work in a light recursive fashion, allowing continuous updates of the phase time series
\item it should provide a quality measure for the estimated phases.
\end{itemize}

\section{Optimizing short-term coherence}
Every time a new acquisition becomes available, it should be possible to generate an interferogram and an interferometric phase. The interferogram will be between a running reference $z_n$ and the new acquisition $y_n$. The estimated phase will be:
\begin{align}
\phi_n & = \angle (\overline z_n \cdot y_n) \label{eq:interferogram}
\end{align}
where the product is understood to be averaged in the spatial domain, for instance over a 2D window. This interferogram will also give a chance to estimate a coherence as a quality measure for the phase $\phi_n$, specific to that acquisition $y_n$ and spatial window. Of course, the core of the problem is defining the running reference $z_n$.

To minimize the phase variance one should maximize the coherence: it makes sense that the reference $z_n$ is a linear combination of the most recent acquisitions, which are the ones with the highest coherence. Explicitly:
\begin{align}
z_n &= {\bf w}^T \: [y_{n-M}e^{-j\phi_{n-M}}, \dots, y_{n-1}e^{-j\phi_{n-1}}]^T \\
&= {\bf w}^T \: {\bf y}_P,
\end{align}
with $\bf{w}$ being a column vector of weights. Notice that the linear combination of past acquisition is possible only after a re-phasing operation. I will assume that ${\bf w}$ is real.

To maximize the coherence one should choose the weights
\begin{align}
{\bf w} = {\bf R}^{-1}{\bf r},
\label{eq:weights}
\end{align}
with ${\bf R} = \text{E}[{\bf y}_P\, {\bf \overline y}_P^T]$ being the variance-covariance of the previous acquisitions, and ${\bf r} = \text{E}[ {\bf y}_P \, \overline y_n]$ the covariance between the previous re-phased acquisitions and the incoming one $y_n$ (but ignoring its interferometric phase, which will be estimated by Eq.~\ref{eq:interferogram}). In this case $z_n$ is the best linear prediction of $y_n$ using only the last $M$ acquisitions, assuming the interferometric phases are known. The appendix shows a derivation of Eq.~\ref{eq:weights} from a coherence-maximization goal.

If we omit the ${\bf R}^{-1}$ equalization, the weights will effectively describe something like an adaptive filter. Moreover, if the correlation $\bf r$ is well approximated by an exponential (within the $M$ samples), it is possible to update recursively the reference $z_n$ in the following way:
\begin{align}
z_{n+1} = \alpha z_{n} + y_n\, e^{-j\phi_n}, \label{eq:recursive}
\end{align}
after having chosen an adequate value for $\alpha$ in the interval $(0,1)$.
This recursive scheme has very minimal storage requirements for a running reference: it would require storing a single (virtual) acquisition and could be updated each time.

\section{Long-term stability}
Updating the running reference $z_n$ is necessary if we want to keep a good short-term coherence. As time passes, the reference would otherwise become obsolete and yield low-coherence interferograms. At the same time, the update process entails the risk of a phase drift. As mentioned before, the drift can result both from stochastic and from systematic effects. To counter the drifts I suggest calibrating periodically the running reference with past versions of itself to guarantee long-term phase stability.

For example one could generate a calibration phase by interfering the current reference with a stable reference $s_n$ and adjust the phase of the running reference:
\begin{align}
\varphi &= \angle (\overline s_n \cdot z_n) \label{eq:calibration_phase}\\
z_n &\leftarrow z_n\, e^{-j \varphi}.
\end{align}
The stable reference $s_n$ could be fixed, for example the first image in the stack ($s_n = y_1$), or it could be a snapshot of the running reference taken after some warm-up time (e.g. $s_n=z_{10}$), or an accumulation of running references:
\begin{align}
s_{n+1} = s_n + z_n. \label{eq:stable_reference}
\end{align}
In any case, after some time, the stable reference should mostly contain components which are coherent in the long run.

A fully recursive algorithms is shown in Alg.~\ref{alg:recursive}. In the pseudo-code, $N$ is the total number of acquisitions.

\algsetup{indent=3em}
\renewcommand{\algorithmiccomment}[1]{// #1}

\begin{algorithm}[h]
\caption[]{Recursive interferometric phase estimation (RIPE)}

\begin{algorithmic}[1]
\STATE \COMMENT{initialization}
\STATE $z \leftarrow y_1$
\STATE $s \leftarrow \alpha \, y_1$
\STATE $\phi_1 = 0$
\STATE \COMMENT{main loop}
\FOR{$n=2$ to $N$}
\STATE $\phi_n = \angle (\overline z \cdot y_n)$ \COMMENT{phase estimation}
\STATE $z \leftarrow \beta z + y_n\, e^{-j\phi_n}$ \COMMENT{running reference update}
\STATE $ \varphi \leftarrow \angle (\overline s \cdot z) $
\STATE $z \leftarrow z\, e^{-j \varphi}$ \COMMENT{running reference adjustment}
\STATE $s \leftarrow s + z$ \COMMENT{stable reference accumulation}
\ENDFOR
\end{algorithmic}

\label{alg:recursive}

\end{algorithm}

\section{Experiments with simulated data}
To compare the performance of the proposed progressive scheme and to a full-covariance method (EMI) we will simulate data according to the following {\it complex} temporal coherence model:
\begin{align}
\gamma(\Delta t) &= 0.18\, e^{-|\Delta t|/11}e^{j 0.03 \Delta t} \nonumber \\
&\quad+ 0.25\, e^{-|\Delta t|/50}e^{j 0.002 \Delta t} +0.13 \nonumber \\
&\quad +0.44\, \delta(\Delta t). \label{eq:model}
\end{align}
The Kronecker $\delta$ is there to guarantee that for zero temporal separation the coherence is 1. The model is stationary and depends only on the time difference $\Delta t$ between any two acquisitions.

This coherence model and its parameters are described in \cite{Ansari2020} and are referred to a particular C-band Sentinel-1 dataset. A peculiarity is represented by the phase terms, introduced to explain phase biases present in short-term interferograms. These phase terms are associated with transient scatterers which are decorrelating exponentially. The stable scatterer component, accounting here for $13\%$ of the total power, is associated with the desired deformation signal.

\begin{figure}[htb]
\centering
\includegraphics[width=9cm]{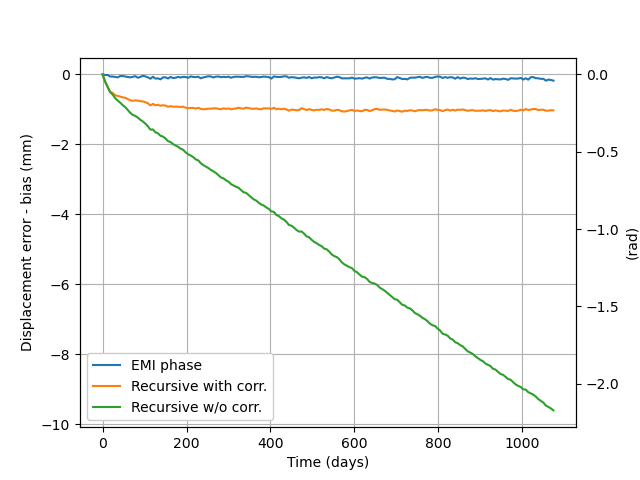}
\caption{The bias of various phase estimators, assuming the coherence model in Eq.~\ref{eq:model}. The phase drift correction (calibration) is clearly necessary.}
\label{fig:bias}
\end{figure}

Figure~\ref{fig:bias} compares the bias of the various algorithms. The bias of EMI is really negligible. The performance of the ``progressive estimation'' is illustrated with two variants, one without any correction of the running reference, the other with a phase drift correction as described by Eq.~\ref{eq:calibration_phase}-\ref{eq:stable_reference}. Both variants are based on a recursive algorithm (Eq.~\ref{eq:recursive}). The variant using a drift control saturates at a bias of about 1~mm, the other shows a drift of about 4 mm/yr. 

The drift control mechanism is clearly necessary; unfortunately it is not effective in the first 100 days or so. This is not really an issue, and can be probably be fixed  retrospectively. It is likely unavoidable given the ``causal'' nature of the algorithm, which has no access to future data, unlike EMI or other full-covariance algorithms.

\begin{figure}[htb]
\centering
\includegraphics[width=9cm]{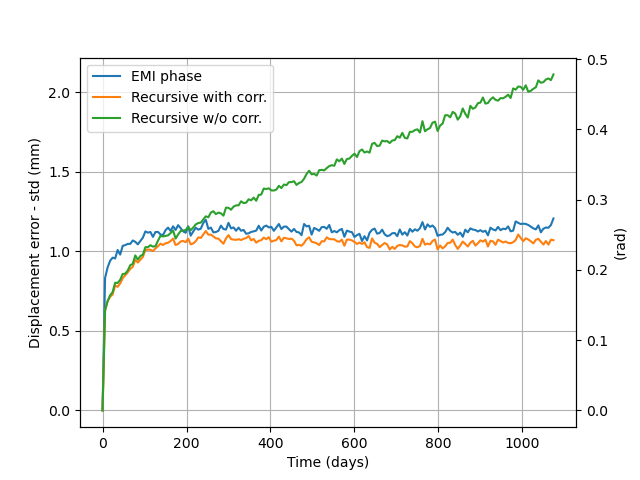}
\caption{The standard deviation of various phase estimators, assuming the coherence model in Eq.~\ref{eq:model}. The drift correction (phase calibration of the running reference) helps limiting the standard deviation in the long term.}
\label{fig:std}
\end{figure}

Figure~\ref{fig:std} reports the standard deviation of $\phi_n - \phi_1$ (the bias is removed in the complex domain to avoid problems with the circularity of the phase).
The discussion on the standard deviation is broadly similar to what could be said for the bias. Without any drift control, the standard deviation increases with time, similar to a random walk. The recursive algorithm with drift control is almost equivalent to EMI. Actually, the short-term variance is smaller for the two progressive schemes than for EMI. This must be related to the imperfect knowledge of the temporal covariance matrix and the necessity of regularization of full-covariance algorithms.

\section{Discussion}
This work shows that a recursive scheme, or more generally a progressive scheme, can reach a performance similar to a full-covariance algorithm like EMI under a realistic coherence scenario. It is therefore possible to produce continuously high-quality phases without having access to the full dataset at the same time, for example when processing during a multi-year mission.

The proposed scheme is explicit in its dealing with different scattering components, whereas full-covariance schemes hide their working behind a global numerical optimization step. The phase estimation is a result of a single interferogram formation (Eq.~\ref{eq:interferogram}) and a calibration procedure (also an interferogram, Eq.~\ref{eq:calibration_phase}). Both steps can be independently designed and controlled, and offer some inspection opportunities, for example by examining the interferogram coherence.

Until now I have discussed the choice of the weights for the running reference only in relation to the prediction of the incoming acquisition. In reality the weights have also an influence on the coherence of the calibration interferogram in Eq.~\ref{eq:calibration_phase}. A running reference that is too skewed towards the short-term coherence will produce a noisy calibration phase for the long term component. One could actually argue that the optimal weights are those that predict simultaneously the incoming acquisition $y_n$ and the stable reference $s_n$ from ${\bf y}_P$. Assuming adequate power normalization for both $y_n$and $z_n$, the optimum weights are given by
\begin{align}
{\bf w} = {\bf R}^{-1}\{ \text{E}[{\bf y}_P \, \overline y_n] + \text{E}[{\bf y}_P \, \overline s_n]\}.
\end{align}
Again, we have to assume a zero-interferometric phase for $y_n$ to write this equation: the actual phase will be estimated later.

In practice, in the recursive algorithm presented above (Alg.~\ref{alg:recursive}), the weights are bound to belong to the exponential family. It is possible to adopt different filter designs for the generation of the running reference (for example, auto-regressive filters of higher order, or ARMA filters) to better approximate the optimal weights. One must also say that the optimal weights will suffer from the usual limitation of imperfect coherence knowledge~\cite{Jiang2020}.

The proposed algorithm filters or estimates only one long-term coherent component ($s_n$). If necessary, it should be possible to filter and keep track of more components to calibrate the running reference. This could be meaningful, e.g., if there is no long-term coherence but a fading coherence with characteristic time of one year. The algorithm may have to be adapted to the actual coherence behavior, which counts as a disadvantage with respect to algorithms based on the temporal coherence matrix.

\section{One experiment with real data}

This experiment was conducted with the same Sentinel-1 dataset as in~\cite{Ansari2020}, a scene taken over Sicily (Italy) from the ascending track with relative orbit number 44. The acquisition period goes from late 2014 until summer 2018. I tested the two variants of the recursive scheme, with and without phase drift corrections. In this case there is no perfect reference to compare with, different from the simulation study. In~\cite{Ansari2020} we have shown that, for this dataset at least, EMI is very close to the phase of Permanent Scatterers; therefore I am taking the EMI result as a reference, assuming its bias is negligible.

\begin{figure}[htb]
\centering
\includegraphics[width=9cm]{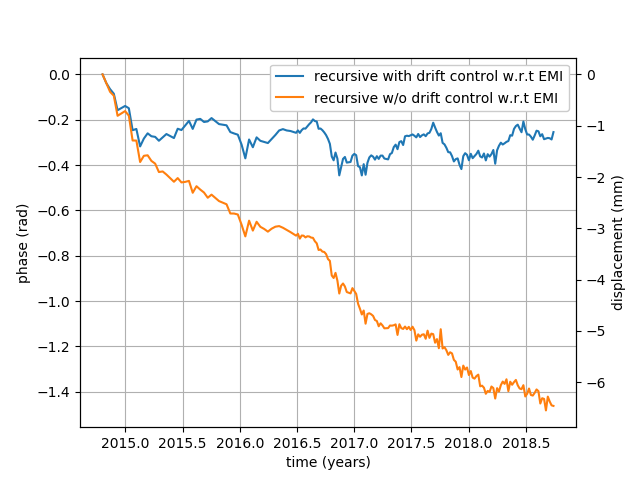}
\caption{The phase bias of the two recursive estimates with respect to EMI (taken as a reference). The need for a drift correction (phase calibration of the running reference) is evident.}
\label{fig:bias_sicily}
\end{figure}

The results in Fig.~\ref{fig:bias_sicily} show that without phase calibration the phase estimation will drift 2-3 mm/yr, with phase calibration the bias is limited between 1-2 mm, oscillating under the influence of some seasonal factor (likely vegetation growth). It is not possible to say whether EMI or the progressive estimation is more affected by the seasonal oscillation, as we observe a difference.

\begin{figure}[htb]
\centering
\includegraphics[width=9cm]{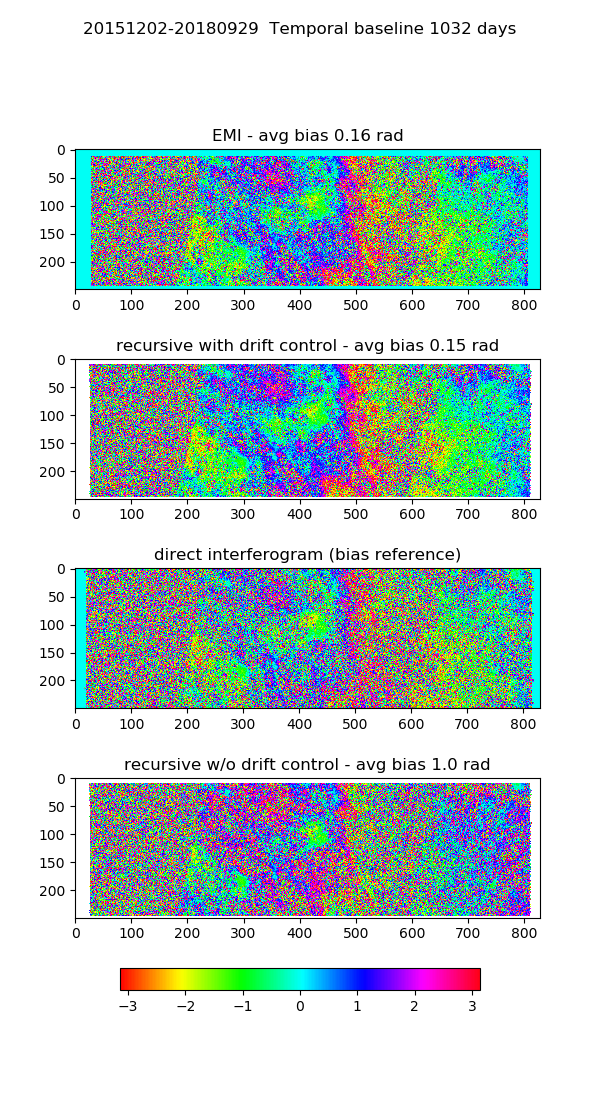}
\caption{A single interferogram estimated by different algorithms for one Sentinel-1 burst. The two images were acquired on 2015-12-02 and 2018-09-29. The direct interferograms is nosier than the other ones. The recursive estimation without drift control yields large biases. The recursive algorithm with drift correction seems to make a good substitute for EMI.}
\label{fig:bias_different_estimates_sicily}
\end{figure}

Figure \ref{fig:bias_different_estimates_sicily} displays the estimated phases for a particular acquisition pair spanning about three years (1032 days), from 2015-12-02 to 2018-09-29. EMI and the recursive estimation are very similar, also in terms of bias compared to the direct interferogram, taken as a bias reference. Without the drift control mechanism, the phase drift is evident to the eye. It is possible also to notice the difference in noisiness in the various estimates.

\begin{figure}[htb]
\centering
\includegraphics[width=9cm,trim=30mm 0mm 35mm 10mm,clip]{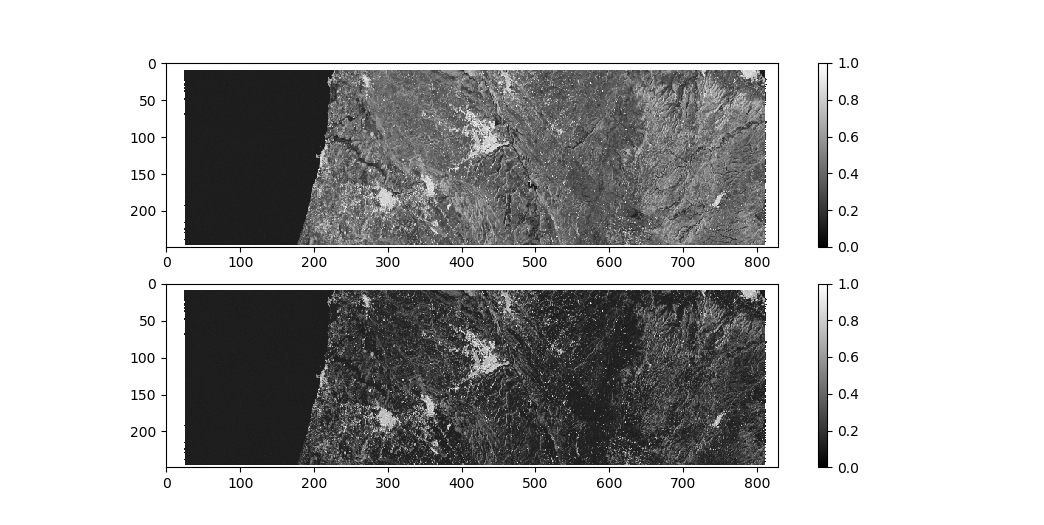}
\caption{Short-term (above) and long-term (below) average coherence magnitudes.}
\label{fig:coherences}
\end{figure}

The short and long-term average coherence magnitudes are displayed in Fig.~\ref{fig:coherences}. They correspond to the interferograms between $y_n$ and the two references ($z_n$ and~$s_n$). The long-term coherence allows discarding targets whose long-term behaviour is not reliable.

\section{Conclusion}
This paper has introduced a phase estimation scheme for InSAR stacks with different characteristics compared to many published algorithms. This scheme does not make use of temporal covariance matrices but builds the phase in a progressive way. A pure recursive implementation has also been presented, which is suited to a continuous update of phase products with minimal buffer requirements. The key point is the adjustment of the interferometric reference which proves beneficial for both systematic and random drifts.

The simulations and a small experiment on real data show the feasibility of this simple approach and its performance close to full-covariance matrix algorithms.

The idea of distilling multiple reference images, corresponding to different scattering components, could be a starting point for further developments.

\section*{Appendix}
This appendix shows that Eq.~\ref{eq:weights} indeed maximizes the coherence of the interferogram in Eq.~\ref{eq:interferogram}.
We want to maximize
\begin{align}
\text E[z_n \cdot \overline y_n] = \text E[{\bf w}^T \: {\bf y}_P \cdot \overline y_n] = {\bf w}^T {\bf r},
\end{align}
subject to the constraint
\begin{align}
\text E[|z_n|^2] = \text E[{\bf w}^T \: {\bf y}_P \cdot {\bf \overline y}_P^T \: {\bf w}] = {\bf w}^T {\bf R} \, {\bf w} = 1.
\end{align}
The constraint guarantees that the variance of $z_n$ is 1, so that the correlation is automatically normalized to a coherence.

The constrained optimization is easily accomplished by maximizing the Lagrangian
\begin{align}
L({\bf w}, \lambda) = {\bf w}^T {\bf r} -\lambda  ({\bf w}^T {\bf R} \, {\bf w} - 1);
\end{align}
which gives
\begin{align}
& {\bf r} -\lambda  {\bf R} {\bf w} = 0 \\
& {\bf w} = \frac{1}{\lambda}{\bf R}^{-1} {\bf r}.
\end{align}
The normalization with $\lambda$ is irrelevant for the phase estimation and we can eventually drop it.

\section*{Acknowledgments}
I have to acknowledge discussions with Ramon Brcic and Simon Zwieback on the relevance of short and long term coherence.
Many thanks to the Copernicus program for the Sentinel-1 data.

\bibliographystyle{IEEEbib}
\bibliography{strings}

\begin{thebibliography}{1}

\bibitem{Teba2008}
A.~Monti Guarnieri and S.~Tebaldini,
\newblock ``On the exploitation of target statistics for {SAR} interferometry
  applications,''
\newblock {\em IEEE Trans. Geosci. Remote Sensing}, vol. 46, pp. 3436--3443,
  Nov 2008.

\bibitem{DeZan2007}
F.~De Zan, F.~Rocca, and A.~Rucci,
\newblock ``{PS} processing with decorrelating targets,''
\newblock {\em Proc. of the Envisat Symposium 2007}, pp. 1--5, July 2007.

\bibitem{Fornaro2015}
G.~{Fornaro}, S.~{Verde}, D.~{Reale}, and A.~{Pauciullo},
\newblock ``{CAESAR}: An approach based on covariance matrix decomposition to
  improve multibaseline–multitemporal interferometric sar processing,''
\newblock {\em IEEE Transactions on Geoscience and Remote Sensing}, vol. 53,
  no. 4, pp. 2050--2065, 2015.

\bibitem{Samiei2016}
S.~{Samiei-Esfahany}, J.~E. {Martins}, F.~{van Leijen}, and R.~F. {Hanssen},
\newblock ``Phase estimation for distributed scatterers in {InSAR} stacks using
  integer least squares estimation,''
\newblock {\em IEEE Transactions on Geoscience and Remote Sensing}, vol. 54,
  no. 10, pp. 5671--5687, 2016.

\bibitem{Ansari2018}
H.~{Ansari}, F.~{De Zan}, and R.~{Bamler},
\newblock ``Efficient phase estimation for interferogram stacks,''
\newblock {\em IEEE Transactions on Geoscience and Remote Sensing}, vol. 56,
  no. 7, pp. 4109--4125, 2018.

\bibitem{Ansari2017}
H.~{Ansari}, F.~{De Zan}, and R.~{Bamler},
\newblock ``Sequential estimator: Toward efficient {InSAR} time series
  analysis,''
\newblock {\em IEEE Transactions on Geoscience and Remote Sensing}, vol. 55,
  no. 10, pp. 5637--5652, 2017.

\bibitem{Ansari2020}
H.~Ansari, F.~De~Zan, and A.~Parizzi,
\newblock ``Study of systematic bias in measuring surface deformation with
  {SAR} interferometry,''
\newblock {\em IEEE Transactions on Geoscience and Remote Sensing}, 2020,
\newblock published online.

\bibitem{Jiang2020}
M.~{Jiang} and A.~{Monti Guarnieri},
\newblock ``Distributed scatterer interferometry with the refinement of
  spatiotemporal coherence,''
\newblock {\em IEEE Transactions on Geoscience and Remote Sensing}, vol. 58,
  no. 6, pp. 3977--3987, 2020.

\end{thebibliography}

\end{document}